# Multi-Tenancy Issues with Service Delivery in Developing Economies: Privacy, Trust and Availability Concerns

**Ezer Osei Yeboah-Boateng, Ph.D.**[#1]**, Akosua Boakyewaa Appiah-Nketiah**[#2]**,**

#1 Faculty of Informatics, Ghana Technology University College (GTUC), Accra, and
eyeboah-boateng@gtuc.edu.gh

#2 Graduate Student, Coventry University, Accra Campus, and loctovia@gmail.com

**ABSTRACT**

Cloud computing is a new paradigm and innovation in the technology service delivery. It is utilized for IT-enabled value creation. The capex-free nature of cloud service delivery renders it very attractive to many SMEs. But it is saddled with multi-tenancy issues; prominent under this study are concerns of privacy, trust and availability. How do end-users trust providers with their sensitive data? How secured and confidential are their corporate assets? Amidst the perennial power outages (a.k.a. "Dumsor"), what is the acceptable available uptime? We sampled and interviewed cloud service providers (CSPs) as well as end-users in Ghana, a developing economy. We also gleaned through some secondary data to ascertain some operational concerns. The results indicate that security and service level agreements (SLAs) are key concerns in respect of privacy and trust issues. Similarly, perennial power outages and security were key availability concerns. This was expected as end-users use cloud services for mission critical information assets, and so requires high availability. The implications are that the cyber-security concerns ought to be addressed if SMEs in developing economies are to adopt and accept cloud computing resources for IT-enabled competitive advantage.

**Corresponding Author:** Dr. Ezer Osei Yeboah-Boateng (eyeboah-boateng@gtuc.edu.gh)

## INTRODUCTION

Computing and Internet usage amongst organizations are deemed as vital tools needed to make business decisions and products deployment [1]. Cloud computing is a new technological paradigm and one of the emerging trends in computing and businesses. Coupled with that, is a business model to save cost and deploy scalable computing resources in fast and efficient manner [2]. With the help of cloud computing, information can now be accessed anywhere irrespective one's country of domicile. The adoption and utilization of this paradigm shift, called Cloud computing, has become pervasive, though Ghana and other developing economies are relatively slower in taking full advantage of the technology [3]
Cloud computing is an IT service delivery model that makes remote resources available for many tenants via a server on a pay-as-you-go basis. At the moment, there is dearth of cloud computing literature on developing economies; so this exploratory study seeks to assess some multi-tenancy issues, such as Trust, Privacy and Availability.

The capex-free nature of cloud service delivery renders it very attractive to many SMEs. In spite of that, it is saddled with multi-tenancy issues; prominent under this study are concerns of trust, privacy and availability. How do end-users trust providers with their sensitive data? How secured and confidential are their corporate assets? Amidst the perennial power outages (a.k.a. "Dumsor"), what is the acceptable available uptime?

This paper is organized as follows: this preamble deals with the background and it's followed by the literature review on cloud computing and associated multi-tenancy issues. The methodology is presented next, followed by the results and analysis, and the implications of those concerns are discussed in the conclusion.

**LITERATURE REVIEW**
*The cloud as a sharing technology is one of the most breathtaking phenomena in the age of the internet as it dissolves physical barriers and opens up new possibilities*. Zabalza et al. [4] defined cloud computing as an information technology (IT) deployment model, based on virtualization, where resources, in terms of infrastructure, applications and data are deployed via the internet as a distributed service by one or several service providers where the services are scalable on demand and can be priced on a pay-per-use basis. Cloud computing systems essentially grant access to large pools of data and computational resources via an assortment of interfaces similar to existing grid and HPC resource management and programming systems [5].

Calheiros et al. [6]stated that "*cloud computing focuses on delivery of reliable, secure, fault-tolerant, sustainable, and scalable infrastructures for hosting internet-based application services.*"

Multi-tenancy involves sharing an application instance among numerous tenants (set of users sharing a general access with particular privileges to that software instance.) usually by providing each tenant a dedicated "share" of the instance, which is cut off from other shares with regard to performance and data privacy [7].

There is no standardized definition of cloud computing, however, the study explain cloud computing and relates it to multi-tenancy as an technology service delivery model that makes remote computing resources available for numerous tenants (multi-tenancy) at a time through a server on a pay-as-you-go basis.

*CLOUD PROSPECTS AND SERVICE DELIVERY*

Generally, there exist three (3) kinds of deployment models: public cloud, private cloud and hybrid cloud.

**Private Clouds**
This type of model doesn't bring much in terms of cost efficiency [8]. It is also called an internal cloud [9]. This can be compared to purchasing, building and maintaining your own infrastructure. Notwithstanding this, it has a huge favor tipping toward security. Security issues in private clouds can be addressed through VPN (secure-access) and even firewall systems. Not surprisingly different countries have different laws and regulations for managing and handling data mostly because of the differences in law jurisdictions which can hinder business if cloud is under different jurisdiction.

**Public Clouds**
Zabalza et al. [4] refer to the it as a standard model of Cloud Computing, where the service is available to anyone on the Internet infrastructure for free or by paying certain amount of money. The infrastructure utilized is usually made accessible to the public. It is mainly owned by cloud service providers (CSP) selling or delivering cloud services. Some popular public cloud services are Amazon EC2 (Elastic Cloud), Google AppEngine and SalesForce.com [9].

**Hybrid Clouds**
Usually, a hybrid cloud is made up of two or more clouds (private, or public).

*Multi-Tenancy Issues*
With the onset of every technology, there are usually opportunities that emanate from them. There are also corresponding challenges that are associated with them, and cloud computing is no exception. Private cloud is most optimized among the models in security as several researches [10] confirm is a very big concern in the clouds especially in multi-tenancy. Users' concerns about security and privacy of data and applications remain a roadblock to cloud adoption. Even though options are available for securing data in the cloud technologically, but many prospective as well as cloud users fall short in giving adequate sincere thought to securing the systems. This is an issue that not only Content Service Providers (CSVs) but all stakeholders must tackle through efforts, further research and development.
The transfer of legacy systems to the cloud has also proved to be problematic because of the expertise available in companies, and also because this act requires thoughtfulness and careful handling in tailoring the application and interfaces as required.
Network Access can become a 'headache' literally. The whole idea of cloud computing relies on the constant nature of network access. Ensuring appropriate high-bandwidth network connectivity is crucial to successfully using the cloud. However, many users as well as Internet service providers overlook the enhancement of their network infrastructure and capacity to handle increased traffic (both from and to the cloud).
To realize the benefits of clouds, an enterprise must develop a comprehensive cloud strategy that also examines and addresses potential risks, including cloud unavailability and failures, as well as compliance requirements, as applicable. Trusting the cloud doesn't eliminate the need for risk management or disaster recovery in a business.

**Multi-Tenancy in the Cloud**
Multi-tenancy falls under a broad umbrella of 'Shared Technology' in Cloud computing. Shared technology can be explained as a combination of operations utilized by multiple entities, such as co-location tenants in a data center. Different researchers have come forth with different and equally similar definitions for multi-tenancy, among which are:
> Afkham[11] explains multi-tenancy as the utilization of the same service or application provided by the CSP regardless of the underlying resources by two or more customers. Subashini&Kavitha[12] also explain multi-tenancy asresourcesharing in Cloud computing while further explaining a resource as anyreusable object in the Cloud infrastructure.

This study employs multi-tenancy as the distribution of cloud resources with emphasis on the delivery of the cloud services amongst sets of concurrent users. Although this also means that

84

the client and any possible attacker will also share the same server (a very likely conducive environment for the attacker). Following [12]'s definition, the reusable objects they mentioned must be cautiously restricted and supervised because they can become the susceptible points of failure within the system and provide access to any intrusion or back door. The issues of multi-tenancy vulnerabilities and possible susceptibilities in respect of trust and privacy are further highlighted in [13]and [14]. They demonstrated how the severity of risk are associated with the layers in the service delivery. These layers are depicted in the data center shown below:

A cloud data center gives us a clear picture of what goes on in the cloud infrastructural environment. A typical data center according to [5] may look like depicted below:

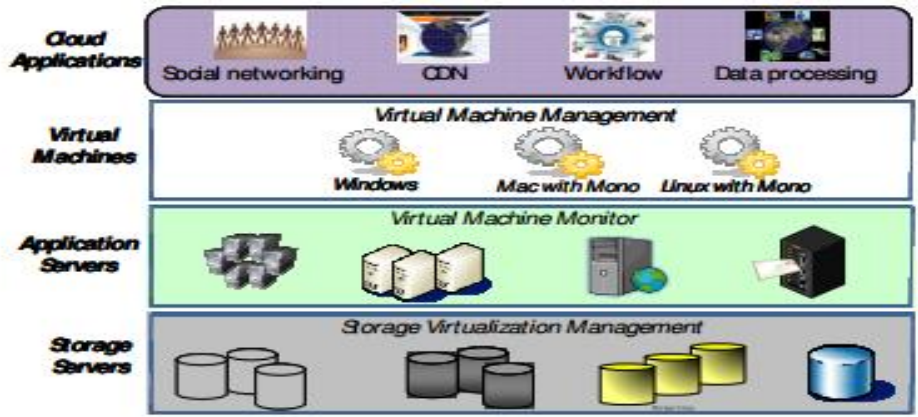

**Figure** Error! No sequence specified.**: data center [5]**

Other issues in multi-tenancy are privacy concerns from the re-usability of resource objects through data surplus (a client with storage access space from aCSP can perform deliberated scans to find sensitive that might have been an unintentional leftover from Virtualization in Multi-tenancy of other customers [13].

One significant challenge in security in Cloud Computing is the trade-off between security and cost. Commenting on the issues, Tim Watson of De Montfort University, notes in [15] as captured in [13, p. 2], that''...although one provider may offer a wonderfully secure service and another may not, if the latter charges half the price, the majority of organizations will opt for it as they have no real way of telling the difference". If customers rely only on price as the main determinant of Cloud service without considering security among other aspects, then security might end up being costly for them.

**Trust Concerns in the Clouds**

There is no universally accepted definition for the complex terminology: trust. It is a mental state comprising the objective to accept vulnerability based upon positive expectations of the intentions or behavior of another" [16] Issues of trust arise from the scenario that customers are not in charge of their cloud resources, therefore have to rely on CSPs to protect their data from any form of security breaches or in the case a security breach actually occurs, still rely on CSPs for law suits, penalties incurred from breach of SLAs, insurance actions. Reputation and security of CSPs can build upon the trust with a client or potential one, stabilize an existing one or invariable decline trust between the entities. Trust can be lost quickly: "a single violation of trust can destroy years of slowly accumulated credibility". [10]

Issues of trust exist and play an important role mainly in the public clouds where users have to rely on their CSPs for their data security but the situation may be slightly different in Private Clouds because security is done on-premises of the user and therefore they may even have a hand in controlling their security measures as it will not be solely relying on the CSPs. Some level of trust may exist between a client and his CSP (a weak type of trust) or sometimes issues may bring about no level of trust (lack of it) at all usually from the client side. When there is no form of trust at all (even the weak type of trust), this means that there will be a non-existent relationship between client and CSP and therefore no delivery of service and no adoption of service either. This just goes to say that some level of trust must exist for cloud service delivery to go smoothly. Transparency of operations must exist for the client to put some level of trust in the CSPs because as stated earlier, client have no control of their information in the cloud (especially on the technical handling of the data and they must therefore trust the CSPs in order to relinquish their sensitive data to them) [10].

When suspicion comes into the picture from lack of transparency, distrust will certainly follow and this will lead to no service solicited and therefore rendered. Security of cloud rendered by a CSP can give room for trust. If inadequate security components exist or CSPs have bad security reputation, clients will be reluctant to patronize their services by putting sensitive data in the clouds. According to [10]: "Trust is key to adoption of Software-as-a-Service (SaaS), and transparency is an important mechanism". Furthermore, trust mechanisms need to be propagated right along the chain of service provision.

**Privacy Concerns**
Privacy is an indispensable human right that includes the right to be left alone [10]. Privacy threats in the cloud may be slightly different because it dissents in relation to the type of cloud scenario, nature of information (confidential or public). This can show the degree to the privacy threat (high or low).

In a multi-tenant environment, privacy should be considered in these paramount areas [17]: Verification and identity management, Access control, Trust management, Security issues in service delivery, Privacy of data and its protection. A person who signs up for cloud services with a CSP therefore reserves the right for his data to be left alone by putting systematic measures in place to verify and manage the identities of persons who access the data, secure the data. The CSP has to build on trust issues with the client by prioritizing privacy.

**Availability of Cloud Data**
This has to deal with reliability of data at all times to authorized clients even in times of data breaches, network issues and any other problem that may arise [17]: The underlying fact is that the client after signing up with a CSP should have access to his data at all times. CSPs therefore have to borne all the costs in maintaining the availability of data. On the other hand, details can be stipulated in the SLAs or any other contract that exist between the two parties and any other third-party.

**Addressing Trust, Privacy and Availability Issues**
Addressing issues of privacy, availability and trust requires effort on both clients and CSPs. Clients should know what constitute confidential data to them to restrict sending those through the CSPs to the cloud. CSPs on the other hand should strive to help new clients who have little ideas in this sector to segregate their data into private and public. Also, SLA's drawn between the two parties should be comprehensive enough with security measures, data availability (guaranteed and or allowed up and downtimes in relation to service), contingencies for data loss, disaster recovery plans, ownership of data and what is allowed to be done with data in terms of third-party organizations. At the end of the day, cloud services usage is a question of tradeoffs between security, privacy, compliance, costs and benefits. [10]

*Service Delivery Frameworks*
Cloud computing like any other technology employs certain frameworks for its adoption. We would briefly explore a theoretical framework adopted for this study: the service jungle.

We explore in detail the "service jungle" theory [18]as the main basis of this research; a theoretical base that does not dwell on a single model but will cut across all the cloud deployment models as well as the service delivery models.

CSPs in the delivery of their services may incur cost increases without corresponding profit margins and returns. The decreasing margin therefore decreases the company's competitive advantage and this may lead to the processes known as the "service jungle".

The service jungle occurs when an organization's competitive position is eroded and leads to the situation where they offer an unfathomable number of services (mainly added services), but cannot charge for it. [18]This framework integrates the individualities of service management, improvement programs and managerial behaviors[18]. CSPs should strive to stay out of the service jungle and into the more lucrative "service garden" that avoids all the setbacks of the service jungle.

**METHODOLOGY**
Research of this type and in this field must be meaningfully carried out with a careful planning process. The researcher employed interviews and secondary data review as the main methodologies to conduct this study.

*Research Design*
A mixed design that combined descriptive and explanatory approach was employed to meet the objectives of the study.

**Population**
The population considered for the study embraces all Organizations in Ghana but will further dwell on the organizations that use ICT services such as data centers, storage services, application services, Internet services etc. The samples are carefully selected to include IT managers, Chief Information Officers (CIO s), Security Functionaries, Corporate Executives and decision makers. This section of the population can also be a distinguishing feature of the notional constructs that this study is determining to explore. Purposive sampling technique was utilized to overcome sample selection bias. However, the section under consideration will not be large enough in relation to the entire population that could have been studied. The two methodologies employed were considered above all others because the research seeks an in-depth understanding into a practice that already exists but is piloted by few organizations within Ghana.

**Interviews**
Interviews are question forms which are completed through dialogue (information is obtained through inquiry and recorded by enumerators) with the respondent [19]They are more costly than questionnaires, but they are better for more complex questions, low literacy or less co-operation. An interview falls within three distinct categories [19]Informal conversational interviews, Semi-structured interviews and Standardized open-ended interviews. Interviews were considered as a methodology for this research because:
 a)     Interviews are handy for attaining an increasing insight and perspective into the topic at hand;
 b)     It will allow the respondents (stakeholders and or IT experts in the selected companies) to explain the subject matter very well which is what is important in this research;
 c)     Interviews are generally useful for gathering concepts quotes and stories.

**Secondary Data**
This can be bluntly put as "second-hand" scrutiny and analysis.Hinds et al.[20]in their explained secondary analysis of qualitative data as the use of on hand data to discover answers to research questions which are different from those inquired about in the original research. *Secondary data itself is and can be supportive in structuring successive primary research and, can grant a baseline with which to weigh your primary data collection results against.*

*Data Analysis*
We would employ the Relative Importance Index as a quantitative measure for the study to gain more insight into the issues discovered by ranking them in terms of critically applied to them by the interviewees. Also, we will analyze the qualitative issues discussed by tabulating, grouping issues, comparing and make meaning out of them.

**FINDINGS AND IMPLICATIONS**
The research has discussed Multi-tenancy issues in cloud computing in the Ghanaian setting. A broad categorization was given to all the data collected using the type of company as basis (i.e. Multinational Companies vs. Small and Medium Enterprises (SME's)). We also compared data from the interviews with our adopted theory: "service jungle" and other secondary data gathered.
**Multinational company** or multinationals is a company that spans countries and international markets physically with branches, subsidiaries, factories etc but are all under the control of the headquarters in the home country. There are many factors we can use to describe **SMEs**. The researcher would prefer to describe SMEs with the number of employees in the organization because inflation in developing economies might render categorization invalid if economic data is used. This study would describe an SME as any company that has an employee head-count of at least two hundred and fifty (250). From data gathered, 25% of the organizations under consideration are Multinationals and the remaining 75% of the Organizations were SMEs.

*Cloud Computing Practices in Corporate Ghana*
From data gathered, we can deduce that the in the delivering of services, 50% of Multi-nationals empower resellers in all their Ghanaian dealings so that it is Ghanaian companies that end up dealing with the clients. The other 50% on other hand takes advantage of their current client-base to deliver their services even though they advertise and head-hunt as well as Brokers or resellers in the industry.

All companies under consideration are unique and cannot really be compared to each other because they all provide different services that make them unique individually, however, we can compare their services (similar ones).

In all 37.5% of all companies under consideration have developed their platform for delivering the services and 62.5% rely on external entities or act as Brokers of the service. Interviewees were asked to explain why some of their services were more costly than that of their foreign counterparts. The study revealed that due to the current power crises in Ghana (Dumsor), 75% of all the companies under consideration had their servers outside of Ghana. This cuts down the cost relatively but there are issues of laws in the host country for the CSPs to consider. The other 25% when asked explained that even though their situation was different (servers are located in Ghana), they still incur extra cost of operations from the fuel they constantly purchase to sustain their in-house generators.

*Multi-Tenancy Concerns with Cloud Computing in Ghana*
One of the main challenge faced by these institutions in delivering the service stems from **lack of Confidence in the System** by the users. Data gathered showed that individuals have no problem in Ghana opening a bank account or buying a new SIM card from any of the telecommunications while giving out their personal data to be registered but are very skeptical about putting their data on the internet and subsequently the cloud. An interviewee gives a scenario of their cloud ticketing system used for an event at the National Theatre of Ghana, he states that at the commencement of the program, it would usually be found out that 80% of the viewers would purchase their ticket at the entrance and only a few would purchase online. He continues by stating that: "Daily transactions that are done in the cloud end up being duplicated on paper for 'safe-keeping' ". It would be quite a while before a proper confidence and trust can be put in the system for Ghanaians to feel safe enough to leave the paper-trail.

The '**Ghanaian Mentality**' as many would call it which simply refers to the negative attitude we have toward the treatment of our data that may end up online as a result of local scenes from "Sakawa" (way of duping for individuals via the internet with their personal details or impersonating someone to do that). This just brings up **issues of privacy of data** that is put up in the cloud. '**Dumsor**' has become a very popular terminology in Ghana from some few years back to describe the power outages and crises currently plunging the nation. This has affected almost all businesses in Ghana and we have had many reactions from downsizing to closing up of many SME's. The Dumsor has affected the budding cloud sector in Ghana as well. From the data gathered, it became known that 75% of the companies under consideration operate servers outside Ghana. This is to cut costs because the one company that operates a server in Ghana has to spend extra on fuel for generators to keep servers running around the clock. However this 75% also encounter issues of **privacy and other Laws** in the countries they host their servers. When the data is hosted in external countries, it cuts down costs for the SME's but has its own issues. The data is now subject to the laws of the host country and also the data takes much to go around and come back to the client.

**Dealing With Challenges Discovered.**
To deal with the challenges identified requires a lot of effort on the companies involved as well as the government of Ghana. If the companies continue to deliver good services and protect the data of clients well, it will gradually minimize the negative mentality and boost the confidence level of clients and prospective ones in the cloud system.

This they can achieve by drawing up good and comprehensive SLAs. The organizations should strive and let potential and existing clients know of their security mechanisms because they are already in place and that should build on the foundation trust of the client.

Also the government has a major role to play in eventually solving or mitigating the power problem to enable companies move their servers in Ghana. This will eventually make sure data is available to the user at all times needed. Policies and laws should be brought up to protect the data of individuals who access the cloud because this would in-turn help the country as a whole.

The bank of Ghana already has a policy that prohibits any banking information of individuals to be kept outside Ghana therefore the SME's that have banks as clients deploy the private cloud to host their data on local servers.

Since Cloud Computing is a new and emerging paradigm in Ghana, it is advised that CSPs work together to build a strong base and platform but also stay in a healthy competition with each other to ensure that Cloud Computing strives well in Ghana.

*Extent of Implications Due Trust, Privacy and Availability*
It was realized from the data collected and analyzed that Privacy of data and Availability of data is both positively related to Trust.

Trust has no physical model for calculation but if a business strives to protect the data privacy for its clients and put active measures to make sure that data is available at all times, it will improve on the low trust that exist between the user and the CSP. Once there is patronage of cloud service from the CSP, there exist some level of trust (weak) between the CSP and his Client and it is the actions and or inactions of the CSP that will build upon this trust or destroy it. [10]

From all data gathered we uncovered that "Whoever provides the data to be used owns it". This just means that, whoever owns the data has absolute control about how it is utilized. But is this really the case in situations where there is third party CSPs involved?

Data gathered shows that the Reseller or Broker CSP forms a higher margin of 62.5% than the independent developer CSPs in Ghana who form 37.5%. These CSPs therefore deal with Third-parties in the client data handling. The SLA however stipulates everything about the data usage, security and availability. Third-party organizations only utilize information based on the term stipulated in their contract and SLA with the said company on behalf of the client. The client should however be made aware that these go on in their data handling.

An implication realized was that if the users are aware of the location of their data (which is usually not the case), they will have a semblance of trust for their CSPs for trusting them with that information. Alternatively when this is not the case, the users should be made comfortable that even if their data is in servers outside of Ghana, CSPs will respect the rights to the privacy of their data in the countries they are hosted in.

All the CSPs discussed had one form of security or another ranging from AES encryption, Military Grid Encryption, SSL certificates, Dedicated VPN (Virtual Private Network) and many more, only 25% of these CSPs had regular third-party audits on their security standards. A good security measure breeds trust, helps protect the privacy of the data and this also makes sure the data is protected and can be made available to the user at all times. Also CSPs should be diligent about the staff they hire and keep them focused in the work they do by fully motivating their staff to since they come into contact with the client's data. According to the service Jungle theory, these other tasks of external audits and Staff Handling and motivation are usually not part of the general cost handling techniques of the organization but these tasks add value to what the CSPs do by invariably increasing the trust clients have in them and this will eventually keep CSPs out of the service jungle.

**Relative Importance Index**

The formula for the Relative Importance Index (RII) utilized is $\frac{\sum w}{a+n}$ ($0 \leq \text{RII} \leq 1$)

Where w = weight given to the issues raised by the respondents; a = the highest weight
n= total number considered

**Table** Error! No sequence specified.**: Ranking of Issues using RII (from the highest to the least)**

| TRUST | RII | PRIVACY | RII | AVAILABILITY | RII |
|---|---|---|---|---|---|
| Security | 0.4 | Security | 0.3 | Power outages ('dumsor') | 0.4 |
| SLA | 0.3 | SLA | 0.3 | Security | 0.2 |
| Third-Party Due Diligence | 0.1 | Remote Server Issues | 0.1 | SLA | 0.1 |
| Disaster recovery techniques | 0.1 | Privacy Laws | 0.1 | Continuity of service | 0.1 |
| Key management | 0.05 | Can CSPs view data? | 0.1 | Unauthorized access | 0.1 |
| Service Transparency | 0.02 | Encryption Management | 0.04 | Server Issues | 0.5 |
| Staff control | 0.01 | External Audits | 0.03 | Legal issues on data | 0.03 |
| Data protection techniques | 0.01 | Data deletion | 0.02 | Data quality | 0.01 |
| Sensitive vs. non-sensitive data | 0.01 | Third-party CSPs | 0.01 | Data Monitoring & Access | 0.01 |

From the table, each of the three categories under trust, privacy and availability were allocated a Relative Importance index of 1 to rank issues discovered and subsequently addressed. It is evident that and even though the indexes are different under the distinct categories, Security and the SLA as well as Power issues are prioritized in all and the issues with the lowest indexes are key management and staff control, data deletion and finally data monitoring and access and legal issues on data respectively.

## CONCLUSION

A new paradigm has emerged in the technology field that is putting all researchers on their toes and goes by the name Cloud Computing. Developing nations like Ghana is trying to keep up with this paradigm so as not to be left out by the developed nations. The study aimed to find out about multi-tenancy issues and concerns of trust, privacy and availability under the giant umbrella of cloud computing.

*Summary of Findings*

This study assessed how the service of Cloud solutions is delivered in Ghana with particular emphasis on multi-tenancy issues in trust, privacy and availability. We sampled eight companies in Ghana that deliver such services and thoroughly examined the delivery in the Ghanaian market against the three issues in multi-tenancy. We further explored literature from various researchers and authors of articles, journals and books and utilized interviews and secondary data review as our methodology.

Throughout this research, we have explained cloud computing as a new paradigm in Ghana in which CSPs host client data at a remote location for him access anywhere with the right credentials. We further went on to find out what multi-tenancy is (we likened this terminology to many tenants living in an apartment complex and sharing resources but also keeping a level of privacy) and looked into issues and implications due to trust, privacy and availability of data.

From the study, we found out that privacy and availability positively impact on trust (an immeasurable component). We explained trust as the total believe that something or someone is safe and reliable; privacy as the non-disturbance of online data by unscrupulous people and availability as having data at one's disposal. We found that good security standard adopted by Ghanaian companies help to protect the privacy and availability of data and these in turn improve on trust between CSP and client. We sampled eight CSPs in Ghana as our sample and even though this may not be a total representative of all CSPs in Ghana, issues uncovered are paramount in all CSPs in Ghana.

The Relative Importance Index used in this study helped us rank the issues uncovered. Security issues and SLA were ranked the highest in Trust and Privacy but 'Power problems' was ranked first in 'availability of data'. This may have been because the study was based in Ghana and at a time where Power issues popularly known as 'Dumsor' is a paramount problem for its citizens.

In conclusion, cloud computing is indeed a new paradigm in Ghana [2] but has a long way to travel in Ghana. Even though the study just dwelled on Service delivery in Multi-tenancy issues in trust, privacy and availability, we found out many general issues that may hold this industry back as well as issues specific to Ghana and suggested ways to deal with them.

As [21] indicated in her book about varying privacy laws in different countries, we found from this study that Ghana is no different and there are no clear law (that individuals are aware of) to protect privacy of data.

Finally, the SLA, we found, is a contract that exist between CSPs and their client and contains all issues of data handling, availability, data usage, services, up and down times and all other issues and component that may arise out of the correspondent between the CSP and the client before, during and even after the contract ends. We again found out that CSPs have disaster recovery plans put in place for data protection and availability to enhance client trust.

*Recommendation for Further Research*

It is recommended for the improvement of service delivery in this sector in the future that a comprehensive study is made into the Ghanaian laws on data privacy and how they are used. This can also be compared to laws in countries that host Ghanaian servers remotely.